\documentclass[english,aps,nofootinbib,superscriptaddress]{revtex4}
\usepackage[T1]{fontenc}
\usepackage{babel}
\usepackage{amssymb,amsmath,graphicx,empheq}
\usepackage[all,poly]{xy}
\usepackage{color}
\newtheorem{definition}{Definition}

\makeatletter
 
 \@ifundefined{textcolor}{}
 {%
   \definecolor{BLACK}{gray}{0}
   \definecolor{WHITE}{gray}{1}
   \definecolor{RED}{rgb}{1,0,0}
   \definecolor{GREEN}{rgb}{0,1,0}
   \definecolor{BLUE}{rgb}{0,0,1}
   \definecolor{CYAN}{cmyk}{1,0,0,0}
   \definecolor{MAGENTA}{cmyk}{0,1,0,0}
   \definecolor{YELLOW}{cmyk}{0,0,1,0}
 }

\@ifundefined{date}{}{\date{}}
\usepackage{multirow}

\makeatother

\begin{document}

\title{Conformal variations and quantum fluctuations in discrete gravity}

\date{\today}

\author{A. Marzuoli}
\affiliation{Dipartimento di Matematica `F. Casorati', Universit\`a  degli Studi
di Pavia, via Ferrata 1, 27100 Pavia, Italy}
\affiliation{INFN, Sezione di Pavia, via Bassi 6, 27100 Pavia, Italy}

\author{D. Merzi}
\affiliation{SISSA, via Bonomea 265, 34136, Trieste, Italy}

\begin{abstract}
After an overview of variational principles for discrete gravity, and on the basis of the approach to conformal transformations in a simplicial PL setting proposed by Luo and Glickenstein, we present at a heuristic level an improved scheme for addressing the gravitational (Euclidean) path integral and geometrodynamics.  
\end{abstract}

\maketitle

\section{Introduction}
On the occasion of this special issue in memory of Mauro Francaviglia we are willing to tie together two topics he was interested in all along his scientific career, namely variational principles in relativistic theories of gravity and the study of the conformal structure of spacetimes.  Interesting applications of the latter topic to relativistic cosmology have been addressed in his last two papers \cite{franca1, franca2}.

The scope outlined above will be achieved here from the perspective of discrete gravity as formulated by Regge in \cite{reg1}. A brief account on variational principles for Regge Calculus is given in Section \ref{sec:regge}.
The issue of conformal transformations within a simplicial PL (piecewise linear) context has been always considered quite problematic.

After the breakthrough constituted by the proof of the Poincar\'e conjecture, a renewed interest in looking at Ricci (and other types of curvature) flows also for polyhedral manifolds has flourished and has involved both relativistic physicists (e.g. \cite{miller}) and geometers.
In particular, Luo, Glickenstein and their collaborators have provided a consistent scheme for dealing with discrete conformal transformations, briefly reviewed in Section \ref{discrete}.

In Section \ref{path}, by resorting to the Euclidean path integral approach to discrete gravity, we argue that it would be possible to express the functional measure in terms of  conformal invariants and conformal weights on a suitably defined ``superspace''.

In Section \ref{fluc} we examine the discrete version of Wheeler--DeWitt equation proposed in \cite{HW1} and suggest how it could be translated into the discrete conformal setting for quantum fluctuations of 3--metrics.

\section{Simplicial spacetimes: Euler--Lagrange variational formulation}
\label{sec:regge}

In \cite{reg1} Regge developed an approach to General Relativity -- the 
\emph{Regge Calculus} --  based on the idea of  modeling spacetimes as simplicial PL manifolds (polyhedra).
The first step consists in  constructing the analogous of the (vacuum) 
(Riemannian) Einstein-Hilbert action in $d$ dimensions
\begin{equation}
S[g]=\frac{1}{16\pi}\int_{M}d^{d}x\sqrt{g}R,
\label{SEv}
\end{equation}
for a compact oriented PL manifold $M$ (of a fixed topological type) with simplices of maximal dimension endowed with a flat, positive definite metric. The Minkowskian case is not addressed here (see \cite{sorkin}) because the quantization scheme 
adopted in the following relies on the Euclidean path integral prescription.

A $d-$dimensional polyhedron is flat everywhere except at $(d-2)-$dimensional simplices, the so called \emph{hinges} (or \emph{bones}) and the notion of 
\emph{curvature} for a polyhedron with assigned edge lengths is introduced through  \emph{deficit angles}.
The deficit angle $\beta_{\kappa}$ at a hinge $\kappa$ is expressed by means of the angles $\alpha_{s}$ between the $(d-1)-$dimensional faces of the $d$-simplices $s\in\mbox{star}(\kappa)$ joining at the hinge $\kappa$ (the star of a simplex $\sigma$ in a simplicial complex $K$ is the union of the simplices in $K$ of which $\sigma$ is a face):
\begin{equation}
\beta_{\kappa}=2\pi-\sum_{s\in star(\kappa)}\alpha_{s}.
\end{equation}
The Regge action for a closed $d-$manifold $M$ reads
\begin{equation}
S_{R}[M,\{\ell\}]=-\frac{1}{8\pi}\sum_{\kappa}A_{\kappa}\beta_{\kappa},
\label{SRv}
\end{equation}
where $A_{\kappa}$ is the $(d-2)-$dimensional volume of the hinge and the sum is taken over all $(d-2)-$simplices. Clearly the action is completely determined by the lengths $\{\ell\}$ of the edges, i.e. $1-$dimensional simplices, \emph{cf.} the appendix for a primer on simplicial geometry.

In order to obtain the field equations analogous to vacuum Einstein equations $G_{\mu\nu}=0$ 
(\emph{Regge equations}) we have to find the configuration of edge lengths that makes the action stationary, i.e. 
\begin{equation}
\frac{\delta S_{R}[M,\{\ell\}]}{\delta\ell}=0,
\end{equation}
where, actually, in this discrete setup, the variation is an ordinary gradient 
\begin{equation}
\frac{\partial S_{R}[M,\{\ell\}]}{\partial\ell_{e}}=0,
\label{Rvar}
\end{equation}
with $e$ varying over the set $E$ of edges in $M$.

In three dimensions the hinges are the edges (sometimes referred to as \emph{links}) and the action reads
\begin{equation}
S_{R}[M^3,\{\ell\}]=-\frac{1}{8\pi}\sum_{e}\ell_{e}\beta_{e}.
\label{SRv3}
\end{equation}

In this case the field equations are, thanks to the Schl\"afli differential identity 
(see  \cite{reg1}, \cite{luo2} and \cite{LJ}),
\begin{equation}
\beta_{e}=0
\label{REv3}
\end{equation}
for all edges $e$, reflecting the local Ricci flatness of each solution to Einstein equations. The Schl\"afli differential identity plays a fundamental role in variational  Regge Calculus: it states that for an Euclidean tetrahedron the differential $1-$forms
$\sum_{e}\ell_{e}d\alpha_{e}$ and $\sum_{e}\alpha_{e}d\ell_{e}$,
where the sums are taken over the edges $e$ of the tetrahedron, are exact. This amounts to to treat the Regge action as depending solely on the edge lengths against variations since
$\sum_{e}\ell_{e}d\alpha_{e}=0$ and $\sum_{e}\alpha_{e}d\ell_{e}= dV$, 
with $V$ the volume of the tetrahedron.

In four dimensions the action has the form
\begin{equation}
S_{R}[M^4,\{\ell\}]=-\frac{1}{8\pi}\sum_{f}A_{f}\beta_{f},
\label{SRv4}
\end{equation}
where $A_{f}$ are the areas of the triangular hinges $f$ and by resorting again to the
Schl\"afli differential identity  for an Euclidean 4-simplex, the Regge action reads 
\begin{equation}
\ell_{e}\sum_{f\supset e} \beta_{f}\mbox{cotan}(\theta_{ef})=0
\label{REv4}
\end{equation}
for each  edge $e\in E$, and where $\theta_{ef}$ is the angle in the triangle $f$ opposite to the edge $e$ (note that this system is overdetermined and locality is lost since
the evaluation of the angles involves the metrics of the simplices which share
the edge of length $\ell$).

If $M^4$ has a non-empty $3d$ boundary $\partial M$, and $f'$ is a boundary triangle, then the contribution to the (extrinsic) curvature is given through the angle between the outer normals of contiguous simplices in $\partial M$ meeting at $\kappa$, namely  
\begin{equation}  
\gamma_{\kappa}=\pi-\sum_{s\in star(\kappa)}\alpha_{s},
\end{equation}
and it can be shown (see \cite{harsor}) that the boundary term is given by (primed quantities refer to boundary components)
\begin{equation}  
S_R[\partial M,\{\ell'\}]=-\frac{1}{8\pi}\sum_{f'}A_{f'}\gamma_{f'}+C(\{\ell'\}),
\end{equation}
where the function $C$ depends on the edge lengths on the boundary, but it is otherwise arbitrary. It is worth noticing that this additional term is necessary in particular to ensure the correct composition law for the transition amplitudes to be considered in the path integral formulation, see Section \ref{path}.

A cosmological constant term can be  introduced in a straightforward way, 
and the Regge action would become 
\begin{equation}
S_{R}[M^4, \Lambda,\{\ell\}]=-\frac{1}{8\pi}\sum_{f}A_{f}\beta_{f}+\frac{\Lambda}{8\pi}\sum_{s}V_{4}(s),
\label{SRcc}
\end{equation}
where $V(s)$ is the volume of the $4-$simplex $s$, so that the second summation represents the total volume of the (compact) polyhedron.

Foundational and mathematical aspects as well as applications of Regge Calculus to classical General Relativity have attracted much attention especially in the 1970-80s, but it was in the early 1980s that such an \emph{ab initio} regularized approach 
was recognized as a quite promising scheme to address quantization of gravity 
(we refer to \cite{witu} for a complete bibliography up to the 1990s, see also
\cite{rewi} and \cite{car1}). A brief account on simplicial quantum
gravity in the Euclidean path integral approach together with
a few  pertinent references is reported in Section \ref{path}.

\section{Discrete conformal transformations}
\label{discrete}
Let us start by summarizing some well known facts about Einstein metrics in the smooth case \cite{besse}.
\begin{definition}[Einstein manifold]
Let $M$ be a compact  $d-$dimensional smooth manifold and let $g$ be a Riemannian metric on $M$. Then $M$ is an Einstein manifold when $g$ and the associated Ricci tensor $Ric(g)$ satisfy the following relation:
\begin{equation}
Ric(g)=\Lambda g,\qquad\qquad\mbox{with}\qquad\qquad \Lambda=\frac{1}{d}\frac{S[g]}{V[g]}
\end{equation}
where $V[g]$ is the volume of the manifold
\begin{equation}
V[g]=\int_{M}d\mu_{g}=\int_{M} d^{d}x\sqrt{g},
\label{volume}
\end{equation}
and $S[g]$ is the \emph{Einstein-Hilbert functional}
\begin{equation}
S[g]=\int_{M}d\mu_{g} R(g)=\int_{M} d^{d}x\sqrt{g}R,
\label{EHf}
\end{equation}
\end{definition}

It is also useful to introduce the normalized version of the Einstein-Hilbert functional:
\begin{equation}
S_{V}[g]=\frac{S[g]}{(V[g])^{\frac{d-2}{d}}}.
\label{VEHf}
\end{equation}

Einstein metrics can be characterized as critical points of the functional  
$S_{V}[g]$, or equivalently as critical points of the functional $S[g]$ in \eqref{EHf}
restricted to the set of metrics having volume normalized to 1.

Following the approach proposed by Luo, Glickenstein and collaborators
\cite{luo1, luo2, gli1,gli2}, in order to construct a discretized version of Einstein manifolds, we define a normalized Regge functional by setting
\begin{equation}
S_{V}[\{\ell\}]=\frac{S_{R}[\{\ell\}]}{(V[\{\ell\}])^{\frac{d-2}{d}}},
\end{equation}
where $S_{R}$ is the Regge action given in \eqref{SRv}.
For a PL manifold $M$ with a fixed triangulation $\mathcal{T}$ the analogue of the space of metrics is the space of edge lengths
$$\mbox{met}(M,\mathcal{T})=\left\{\{\ell_{e}\}\in\mathbb{R}_{+}^{|E|}\ \mbox{s.t.}\ (-1)^{d+1}\mbox{CM}_{d}>0\ \mbox{for all simplices in $\mathcal{T}$}\right\}$$
where $E$ is the set of the edges of the triangulation and CM$_{d}$ is the $d-$dimensional Cayley-Menger determinant \eqref{CM}. 

From now on, being $S_{R}[\{\ell\}]$ a \emph{function} depending on the edge lengths, we will denote by $S_{R}(\vec{\ell})$ the function defined on met$(M,\mathcal{T})$. Moreover we will freely refer to these vectors of edge lengths as \emph{metrics}. 

In the rest of this section we will restrict  to $3-$dimensional triangulated manifolds
(still denoted simply by $M$) for which the Regge functional has the form 
\begin{equation}
S_{V}(\vec{\ell})=\frac{S_{R}(\vec{\ell})}{(V(\vec{\ell}))^{\frac{1}{3}}}=\frac{\sum_{e}\ell_{e}\beta_{e}}{(\sum_{\tau}V_{3}(\tau))^{\frac{1}{3}}},
\label{svgli}
\end{equation}
where the sum in the denominator is over all the tetrahedra $\tau$ (note that we dropped the irrelevant factor $1/8\pi$ in the above definition of $S_{R}$).  

Einstein metrics in the triangulated setup are introduced as follows \cite{gli2}:
\begin{definition}[Discrete Einstein metrics]
\label{eindef}
Let $(M,\mathcal{T},\vec{\ell})$ a triangulated 3-manifold. The set of edge lengths $\{\ell\}$ is an \emph{Einstein metric} if it is a stationary point of the normalized Regge functional ${S_{V}(\vec{\ell})}$, namely if
\begin{equation}
\frac{\partial S_{V}(\vec{\ell})}{\partial\ell_{e}}=0
\label{critsv}
\end{equation}
for all edges $e$.
\end{definition}

We can calculate explicitly the variation (\cite{gli2})
\begin{equation}
\frac{\partial S_{V}}{\partial\ell_{e}}= V^{-\frac{1}{3}}\frac{\partial S_{R}}{\partial\ell_{e}}-\frac{1}{3}V^{-\frac{4}{3}}\frac{\partial V}{\partial\ell_{e}} S_{R}=V^{-\frac{1}{3}}\left(\beta_{e}-\frac{1}{3}\frac{S_{R}}{V}\frac{\partial V}{\partial\ell_{e}}\right)
\end{equation}
where we have used the Schl\"afli differential identity. Then the conditions \eqref{critsv} can be written in terms of the \emph{edge curvature} $K_{e}:=\beta_{e}\ell_{e}$ as
\begin{equation}
\label{reggemet}
K_{e} =\Lambda\ell_{e}\frac{\partial V}{\partial\ell_{e}},\qquad\qquad\mbox{where}\qquad\qquad\Lambda=\frac{S_{R}}{3V}.
\end{equation}

The stationary configurations of the normalized functional $S_{V}$ are the same as those of the Regge functional with a cosmological constant term, i.e.  
$S_{\Lambda}=S_{R}-\Lambda V$ (as happens in the smooth case). Indeed the critical points satisfy
\begin{equation}
K_{e}=\Lambda\ell_{e} \frac{\partial V}{\partial\ell_{e}},
\end{equation}
and the value of $\Lambda$ can be found by summing over the edges $e$
\begin{equation}
\Lambda= S_{R}\left(\sum_{e}\ell_{e}\frac{\partial V}{\partial\ell_{e}}\right)^{-1}=\frac{S_{R}}{3V}.
\end{equation}

Coming to the main point of this section, namely the introduction of the notions of \emph{conformal transformation} and \emph{conformal structure} into the triangulated context, a naive attempt might be to take an infinitesimal approach, i.e. asking for discrete conformal transformations, whatever they are, to preserve infinitesimally the angles in analogy with the smooth case. However there is no way to come up with any reasonable definition from this hypothesis, essentially  because ``locality''
cannot be conserved consistently: in every simplex the geometry is Euclidean and the non trivial features of  Regge--like triangulations are encoded into restricted subsets, e.g. the curvature is concentrated at the $(d-2)-$dimensional simplices.\\

The proposed consistent solution  relies on the choice of a local
 point of view,  in analogy with the smooth case in which the conformal class of a metric is defined through the action of positive smooth functions on smooth metrics (i.e. $\tilde{g}(x)=\Omega(x)g(x)$ with $\Omega(x)$ positive in every point $x$). Thus a discrete conformal class of a discrete metric (represented by a vector of edge lengths) is the collection of the discrete metrics related by the action of positive functions defined on the \emph{vertices of the triangulation}. The inherent discrete conformal structures defined below are \emph{triangulation dependent}. 

 \begin{definition}[Discrete conformal structure]
 Let $(M,\mathcal{T})$ be a triangulated manifold and let $U$ be an open subset of $V^{*}(\mathcal{T})$  defined according to
$$V^{*}(\mathcal{T})=\left\{\mathfrak{f}:
V\rightarrow\mathbb{R}\right\}\simeq\mathbb{R}^{|V|}$$ 
where $V(\mathcal{T})$ is the set of the vertices of the triangulation $\mathcal{T}$
(the symbol $V$ referred to vertices is not to be confused with the the
same symbol used for volumes).\\
A \emph{conformal structure} $\mathcal{C}(M,\mathcal{T},U)$ is a smooth map
$\mathcal{C}(M,\mathcal{T},U):U\rightarrow\mbox{\emph{met}}(M,\mathcal{T})$
such that if $\vec{d}=\mathcal{C}(M,\mathcal{T},U)[\mathfrak{f}]$ then for each edge $e$ labeled by $(ij)$ and for every vertex $v$ labeled by $k$:
\begin{equation}
\frac{\partial\ell_{ij}}{\partial \mathfrak{f}_{k}}=\delta_{ik}d_{ij} + \delta_{jk}d_{ji}
\end{equation}
with $\ell_{ij}=d_{ij}+d_{ji}$ and where $\delta$ is the Kronecker delta.\\
A \emph{conformal variation} of a triangulated manifold with fixed edge lengths $\vec{D}$ is a smooth curve
$$\mathfrak{f}:(-\varepsilon,\varepsilon)\rightarrow V^{*}(\mathcal{T})$$
such that there exist a conformal structure $\mathcal{C}(M,\mathcal{T},U)$ with $\mathfrak{f}(-\varepsilon,\varepsilon)\subset U$ and $\mathfrak{f}(0)=\vec{D}$. We call such a conformal structure an \emph{extension} of the conformal variation.
\end{definition}

In the following we will use only a particular conformal structure, called \emph{perpendicular bisector conformal structures} (\cite{gli2}) and we will refer to it simply as conformal structure.

\begin{definition}[Perpendicular bisector conformal structure]
\label{confgli}
Let $(M,\mathcal{T},L)$ be a triangulated manifold with fixed edge lengths $\vec{L}$ and let $U$ be an open subset of $V^{*}(\mathcal{T})$. A \emph{perpendicular bisector conformal structure} $\mathcal{C}(M,\mathcal{T},U)$ is a smooth map
$\mathcal{C}(M,\mathcal{T},U):U\rightarrow\mbox{\emph{met}}(M,\mathcal{T})$ 
determined by
\begin{equation}\label{confl}
\ell_{ij}=\exp\left[\frac{1}{2}(\mathfrak{f}_{i}+\mathfrak{f}_{j})\right]L_{ij},
\end{equation}
where $i,j$ are vertices labels. The \emph{conformal class} is the image of $U$ in $\mbox{met}(M,\mathcal{T})$, and it is completely determined by $\vec{L}$.
A \emph{conformal variation}  $\mathfrak{f}(t)$ is a smooth curve
$$\mathfrak{f}:(-\varepsilon,\varepsilon)\rightarrow U,$$
and it induce a \emph{conformal variation of metrics} $\vec{\ell}(\mathfrak{f}(t))$.
\end{definition}

For a \emph{fixed conformal class} and for any tetrahedron (whose vertices are labeled by 1, 2, 3, 4, according to Fig. \ref{labelt}) in the triangulation there exist two constant $\xi$ and $\eta$:
\begin{equation}
\label{xieta}
\xi=\frac{\ell_{12}\ell_{34}}{\ell_{14}\ell_{23}},\qquad\qquad\eta=\frac{\ell_{12}\ell_{34}}{\ell_{13}\ell_{24}},
\end{equation}
which can be interpreted as ``conformal invariants'' (see \cite{ssp},\cite{gli2}).\\ 
Coming now to the notion of curvature concentrated at vertices, note preliminarily that it must depend on the choice of conformal structure. 

\begin{definition}[Vertex curvature]
\label{vertcurv}
The \emph{vertex curvature} associated with  a vertex $v$ is
\begin{equation}
K_{v}=\frac{1}{2}\sum_{e\supset v}K_{e},
\end{equation}
where the sum is taken over all edges $e$ containing $v$.
\end{definition}
  
The $3d$ Regge action can be written in term of $K_{v}$ according to
\begin{equation}
S_{R}(M,\mathcal{T},\vec{\ell})=\sum_{v}K_{v}.
\end{equation}

Consider conformal variations of $S_{R}$ and $S_{V}$: for a conformal variation 
$\mathfrak{f}(t)$ we have
\begin{equation}
\frac{\partial}{\partial \mathfrak{f}_{v}}S_{R}(M,\mathcal{T},\vec{\ell}(f))=K_{v},
\end{equation}
\begin{equation}
\frac{\partial}{\partial \mathfrak{f}_{v}}S_{V}(M,\mathcal{T},\vec{\ell}
(\mathfrak{f}))=V^{\frac{1}{3}}\left(K_{v}-\frac{S_{R}}{3V}V_{v}\right),
\end{equation}
with
\begin{equation}
\label{Vvert}
V_{v}=\frac{1}{3}\sum_{\tau\supset f\supset v}A_{f}h_{f\tau}
\end{equation}
where $A_{f}$ is the area of the face $f$ and $h_{f\tau}$ is the distance between the face $f$ and the circumcenter of the tetrahedron $\tau$. Upon summation of $V_{v}$ over all $v$, on each tetrahedron we count three times every sub-tetrahedron, so that
$$\sum_{v}V_{v}=3V, \qquad V_{v}=\frac{\partial V}{\partial f_{v}}.$$

This vertex--based objects can be used to define a discrete analogue of constant
curvature metrics: 
\begin{definition}[Constant curvature discrete metrics]
\label{cscm}
A $3d$ triangulation $\mathcal{T}$ has \emph{constant scalar curvature} if it is a critical point of $S_{V}$ with respect to conformal variations 
\begin{equation}
K_{v}=\frac{S_{R}}{3V}V_{v}
\end{equation}
\end{definition}
It can be shown that the set of Einstein metrics is a subset of the set {constant scalar curvature metrics \cite{gli1}.

We now consider a particularly simple triangulation of the 3-sphere $\mathbb{S}^{3}$:
\begin{definition}[The Double Tetrahedron (DT)]
The Double Tetrahedron  is the triangulation of $\mathbb{S}^{3}$ obtained by identifying the boundary faces of two disjoint tetrahedra. In particular this triangulation is characterized by 4 vertices, 6 edges, 4 triangles, and 2 tetrahedra.    
\end{definition}
Note that  this triangulation is ``singular'' in the sense that it cannot be embedded isometrically in $\mathbb{R}^{4}$. This feature can be somehow visualized 
in the $2d$ analogue, i.e.  $\mathbb{S}^{2}$ in $\mathbb{R}^{3}$ triangulated with two triangles whose vertices and edges are identified as shown  in Fig. \ref{pillow}
\begin{figure}[htbp]
\begin{center}
\includegraphics[scale=0.2]{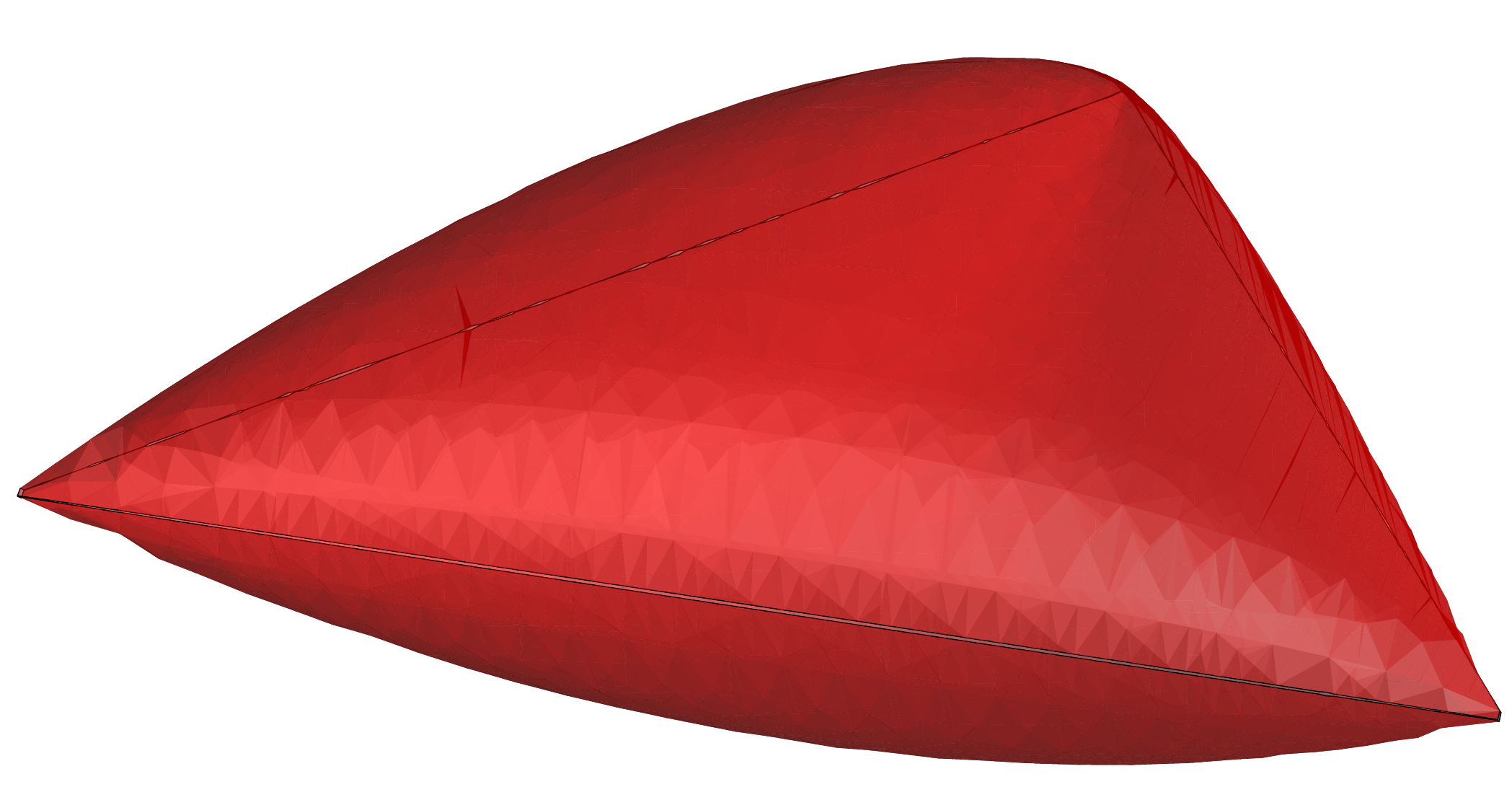}
\caption{\small{The ``Double Triangle'' triangulation of $\mathbb{S}^{2}$} }
\label{pillow}
\end{center}
\end{figure}

The particular identifications of the Double Tetrahedron give it a very simple structure: in fact, being the edges pairwise identified, the two tetrahedra are forced to have the same volume and the same dihedral angles $\alpha_{e}$.
Then  the Regge action is recasted into
\begin{equation}
S_{R}^{DT}=\sum_{e\in E}K_{e}=\sum_{e\in E}(2\pi-2\alpha_{e})\ell_{e},
\end{equation}
and the volume is simply two times the volume $V_3(\tau)$ of a single tetrahedron $\tau$, i.e.
\begin{equation}
V_{DT}=2V_{3}(\tau)=2\sqrt{\frac{CM_{3}}{288}}=\sqrt{\frac{CM_{3}}{72}}.
\end{equation}

The space of metrics (i.e. edge lengths) for the Double Tetrahedron is
\begin{equation}
\mbox{met}(DT)=\left\{\vec{\ell}\in\mathbb{R}_{+}^{6}: CM_{3}>0\right\}.
\end{equation}

Let us consider now a very particular class of metrics, the \emph{equal length metrics}. It is easy to verify that they are stationary points the normalized functional $S_{V}$: in fact, for a Double Tetrahedron whose edges are all long $L$ (we indicate this metric with $\vec{L}$), the Regge action, the volume and its derivative have the following forms
\begin{align}
S_{R}^{DT}(\vec{L})=6K_{e}(\vec{L})=2L\left(\pi-\arccos\left(\frac{1}{3}\right)\right),\\
V_{DT}(\vec{L})=\frac{1}{3\sqrt{2}}L^{3},\qquad\qquad \frac{\partial V_{DT}}{\partial \ell_{e}}(\vec{L})= \frac{1}{6\sqrt{2}}L^{2}.
\end{align}
From the above relations it can be easily shown that $\vec{L}$ satisfies \eqref{reggemet} and therefore equal length metrics are Einstein metrics in the sense of Definition \ref{eindef}. Moreover, evaluating the Hessian of $S_{V}$ at the points $\vec{L}$, we find  that they are \emph{local minima} since  the Hessian has positive eigenvalues, as shown in the following table 
\cite{gli2},

\begin{table}[htbp]
\centering
\begin{tabular}[c]{ |c|c|c | }
\hline
Eigenvalues & Eigenspaces & Spanning vectors   \\ 
\hline
\hline
0&$V_{0}$&$(1,1,1,1,1,1)$\\ 
\hline
\multirow{3}{*}{$\lambda_{1}=2^{\frac{7}{6}}3^{-\frac{2}{3}}(2^{\frac{3}{2}}+9\pi-9\arccos(\frac{1}{3}))L^{-2}$}&{}&$(1,0,0,0,0,-1)$\\ &$V_{1}$ & $(0,1,0,0,-1,0)$\\ &{} & $ (0,0,1,-1,0,0)$\\
\hline
\multirow{2}{*}{$\lambda_{2}=2^{\frac{7}{6}}3^{\frac{1}{3}}(-2^{\frac{3}{2}}+7\pi-7\arccos(\frac{1}{3}))L^{-2}$}&$V_{2}$&$(0,1,-1,-1,1,0)$\\ & {} & $(1,-\frac{1}{2},-\frac{1}{2},-\frac{1}{2},-\frac{1}{2},1)$\\
\hline
\end{tabular}
\end{table}

The 0 eigenvector is clearly related to \emph{scaling}: it relates equal length metrics, which are obtained one from the other exactly by a uniform rescaling of the edge lengths. In general scaling direction is someway irrelevant since the action is constant along it: a uniform rescaling does not change the action (this is true even if we consider a non-stationary metric).
It is still to determine if the equal length metrics are the only minima of $S_{V}^{DT}$ and, moreover, since the Hessian is not globally positive definite we cannot conclude that they are global minima.\\  
Turning  to conformal classes, we have that for the Double Tetrahedron  the two constant $\xi$ and $\eta$  identify completely the conformal class, i.e. the couple $(\xi,\eta)$ in Eq. \eqref{xieta} can be used to parametrize conformal classes of metrics on DT.
Then the structure of met$(\mathbb{S}^{3},DT)$, in brief met$(DT)$, is the following: the space is an open subset of $\mathbb{R}^{6}$, that can be parametrized (in a non-unique way) by $\xi$, $\eta$ and the four conformal factors $\mathfrak{f}_{k}$
of the vertices, see Def. \ref{confgli}. Upon variation of $(\xi,\eta)$ we obtain 4-dimensional submanifolds of met$(DT)$ corresponding to conformal classes.

In the attempt to find \emph{constant scalar curvature metrics} on the Double Tetrahedron a key role is played by \emph{equihedral tetrahedra}:
\begin{definition}[Equihedral tetrahedra]
An \emph{equihedral tetrahedron} is a tetrahedron such that all its faces are congruent, or, equivalently,  such that opposite edges have equal lengths.
\end{definition}


An interesting property of equihedral tetrahedra is that all of the volumes entering  the definition of $V_{v}$ in Eq. \eqref{Vvert} are equals. 
It can be shown (see \cite{gli2}) that on $DT$ \emph{equihedral metrics} (i.e. metrics such that all the tetrahedra are equihedral) have \emph{constant scalar curvature} in the sense of Definition \ref{cscm}. Moreover there is a \emph{unique} equihedral metric \emph{up to scaling} in every conformal class.
Note however that it is still unknown it equihedral metrics are the only constant curvature metrics on $DT$, and therefore if constant scalar curvature metrics are unique within a conformal class.\\
Anyway, the uniqueness (up to scaling) of equihedral metrics within a conformal class reveals a sort of ``transversality'' between the submanifold of equihedral metrics and the submanifolds of \emph{conformally related metrics}. The following 
figures are aimed to clarify pictorially such a  conformal landscape.

\begin{figure}[h!]
\begin{minipage}[c]{0.57\textwidth}
\includegraphics[scale=.35]{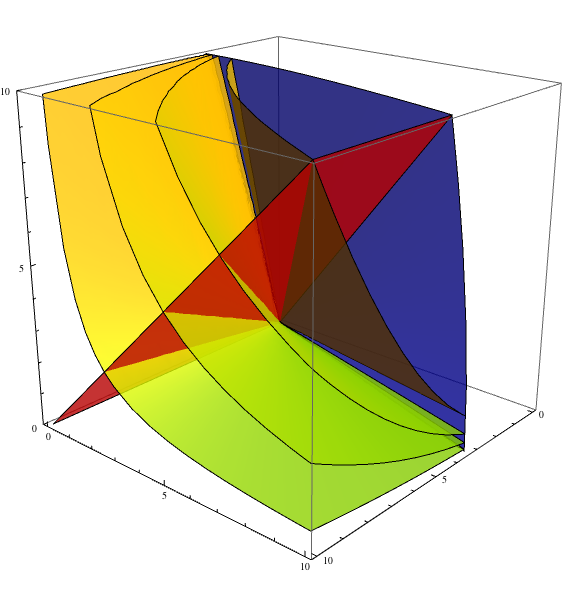}
\end{minipage}
\begin{minipage}[c]{0.4\textwidth}
\caption{Pictorial representation in $\mathbb{R}^{3}$ of the space of metrics of $DT$. This low dimensional picture exhibits the main features described above: in blue we have the boundary given by the constraint $CM_{3}=0$, the yellow ``leaves'' are the conformal classes, and the red plane is the subspace of equihedral metrics. Notice that the uniqueness up to scaling of equihedral metrics within conformal classes is faithfully depicted.}\label{smpic}
\end{minipage}
\end{figure}

\section{The Euclidean path integral approach revised}
\label{path}
Regge discretizations can be used directly to  construct a lattice regularized path integral approach known as \emph{simplicial quantum gravity} and introduced in the 
1980s \cite{rw, hr1, hr2, hr3, mini1, mini2, mini3}.
Here we adopt the formalism and the notations as established
in Hamber's book \cite{ham1}.
 
Recall from section 2 that the role of the metric $g$ is played 
(in any dimension $d$) by the collection of the edge lengths $\{\ell\}$ 
of the polyhedron. The basic ingredient of the simplicial path integral approach is the discrete version of the DeWitt supermetric defined on the space of 
$PL$ simplicial  dissections of a Riemannian $d$-manifold of fixed topology. 
It reads 

\begin{equation}
G_{\scriptsize{DW}}^{ijkl}[g(s)]=\frac{1}{2}\sqrt{g(s)}\left[g^{ik}(s)g^{jl}(s)+g^{il}(s)g^{jk}(s)+\gamma g^{ij}(s)g^{kl}(s)\right],
\end{equation}
with  $\gamma\neq-\frac{2}{d}$ a parameter. Then it can be proven that
\begin{equation}
d[\{\ell\}]=\prod_{s}\left[\det G[g(s)]\right]^{\frac{1}{2}}\prod_{i\geq j}dg_{ij}(s),
\end{equation}
where the first product is taken over all the $d-$simplices $s$, and
$\det G[g(s)]\propto\left(1+\tfrac{1}{2}d\lambda\right)$ $[g(s)]^{\frac{1}{4}(d-4)(d+1)}$.

In the $d=4$ case the measure takes on the particularly simple form
\begin{equation}
\label{m4d}
d[\{\ell\}]=\prod_{s}\Theta(s)\prod_{e}d\ell^{2}_{e},
\end{equation}
where the first product is over $d-$simplices and $\Theta(s)$ is a ``step'' function that takes into account the triangular inequalities and their higher dimensional analogues through the Cayley-Menger determinants (see the appendix)
\begin{equation}
\Theta(s)=\left\{\begin{array}{ll}
1 & \mbox{if the edges of $s$ satisfy $(-1)^{n+1}$CM$_{n}>0$ for all $n\leq d$}\\
0 & \mbox{otherwise}
\end{array}\right..
\end{equation}
The generating  functional  in this discretized setup has the formal expression
\begin{equation}
\label{ZRM}
Z_{R}[M^4, \Lambda]=\int d[\{\ell\}]\mbox{e}^{-S_{R}[M, \Lambda,\{\ell\}]},
\end{equation}
where $S_{R}[M^4, \Lambda,\{\ell\}]$ is the Regge action with a cosmological constant as in \eqref{SRcc}.
This regularized version of the $Z$ functional simplifies in principle the calculations by reducing the functional integral over the space of smooth metrics  to an ordinary (multiple) integral over an Euclidean space (i.e. $\sim\mathbb{R}^{|E|}$ where $|E|$ is the number of edges of the polyhedron). We refer  to the references listed above for a comprehensive account of solved and still open questions
regarding this approach (e.g. based on the addition  of higher--curvature  extra terms,  see \cite{hr1}, \cite{hr2} and \cite{hr3}).\\

To our knowledge the approach of conformal variations in the PL context 
reported in Section \ref{discrete} has never been exploited before in connection with 
the study of quantum amplitudes in the (Euclidean) path integral approach 
to gravity. Here we are going to discuss just a few foundational issues and possible 
advantages of this mathematical background highlighting the differences 
with the standard formalism outlined above. A preliminary attempt of looking at
a simple quantum transition amplitude (topologically associated with 
 a cylinder $\mathbb{S}^3 \times I$) has been addressed in \cite{dm}.
 
 Recall first of all that the scheme is triangulation dependent so that the usual 
 prescription ``summing over triangulations of a manifold of fixed topological type''
 (and eventually going through the continuum limit) does not apply here. The proper arena is met$(M, \mathcal{T})$ for a fixed $\mathcal{T}$, a space which encodes
automatically the prescribed triangular inequalities (the non--local
conditions on the  Cayley--Menger determinants have to be required in 
the standard DeWitt-type measure, see in particular \eqref{m4d}). Clearly 
triangulations with few vertices (and edges) are  more suitable 
to be handled explicitly: met$(M, \mathcal{T})$ is an open subset of
$\mathbb{R}^{|E|}$, the number of (independent) discrete conformal transformations
must be proportional to the number of vertices  and any such a ``superspace'' 
turns out to be bounded by limiting submanifolds $\subset \mathbb{R}^{|E|}$ (of suitable codimensions) where $CM_{n}=0$, $n\leq d$. Of course this picture
is reminiscent of what happens for the simplest triangulation of the $3-$sphere,
the Double Tetrahedron addressed in section \ref{discrete} (to be discussed again in the
next section). It is still an open problem how to extend the mathematical 
setting by Luo and Glickenstein to polyhedra of higher dimension. In our opinion 
the most economical requirement is to take for granted the expression for
$\ell_{ij}$ in \eqref{confl} and write down the Regge action in
$d=4$ by expressing the areas of the triangular hinges in terms of the squares 
of these edge lengths by Heron formula. Of course this hint should be 
applied to concrete few-vertices triangulations, e.g. of the 4-sphere,
in order to make explicit a concrete proposal for the measure on 
met$(\mathbb{S}^4, \mathcal{T})$. An equal lengths Einstein metric 
(denoted $\vec{L}$ in the case of the
Double Tetrahedron) is the best candidate to play the role of
representative in each conformal class, up to an uniform rescaling. 
Then the full measure $d[\{\ell\}]$ will turn out to be parametrized and hopefully 
factorized in terms of the conformal weights of the $N_0$ vertices, 
on the one hand, and of a small number of other parameters like those 
defined in \eqref{xieta} (1 to 3 for the simplest triangulations of the $3-$sphere), 
on the other.
Work is in progress on this directions.

\section{Conformal quantum fluctuations: a case study}
\label{fluc}

Hamber and Williams have recently proposed an improved canonical formulation of simplicial quantum gravity complementary to the Euclidean lattice path--integral discussed in the previous section much on the basis of their previous works. 

A major breakthrough is represented by the explicit expression of the discrete Wheeler--DeWitt equation \cite{HW1,HW2, HW3} - reflecting the existence of the Hamiltonian constraint in the $3+1$ ADM formulation of General Relativity. As is well known a physical vacuum state of a $3-$geometry $\left|\Psi\right>$ in the source--free case has to satisfy $\hat{H}\left|\Psi\right>=0$, and in particular $\left|\Psi\right>$ is to be identified in the position representation as a functional of the $3-$metric $g_{ij}(x)$, denoted $\Psi[g_{ij}(x)]$.

After a careful analysis (see Section 5 of \cite{HW1}), the following ``quite local'' expression of the Wheeler--DeWitt constraint for the wave functional referred to a single tetrahedron $\tau$, $\Psi[\{\ell\}]$ is derived and reads
\begin{equation}
\label{dWdW}
\left\{-(16\pi G)^{2}\sum_{i,j\in \tau}\frac{G_{ij}(\tau)}{4\ell_{i}\ell_{j}}\frac{\partial^{2}}{\partial \ell_{i}\partial \ell_{j}} -2n_{\tau h}\sum_{h\subset \tau}\ell_{h}\beta_{h}+2\Lambda V_{3}(\tau)\right\}\Psi[\{\ell\}]=0 ,
\end{equation}
where the Newton's constant $G$ has been restored and $n_{\tau h}^{-1}$ is the number of tetrahedra which share the hinge $h$. Here $G_{ij}(\tau)$ is the inverse of
\begin{equation}
G^{ij}(\tau)=-\frac{6}{4\ell_{i}\ell_{j}V_{3}(\tau)}\frac{\partial^{2}V_{3}^{2}(\tau)}{\partial\ell_{i}\partial\ell_{i}}.
\end{equation}

If the discrete Wheeler--DeWitt equation is now defined on a space of edge lengths, denoted met in Section \ref{discrete}, the locality requirement can be improved since the Regge curvature and the volume have been introduced as associated to the vertices of any given triangulation (cfr. Def. \ref{vertcurv}). Thus in this setting, where conformal transformations are performed with respect to vertices (and the edges are decorated consistently with \eqref{confl}), the notion of locality emerges more effectively than in the purely simplicial approach.

On the other hand, interpreting \eqref{dWdW} in a superspace $\mbox{met}(M^{3},\mathcal{T})$ sets the stage for a novel implementation of geometrodynamics in the discrete setting.

As a case study consider again the Double Tetrahedron $DT$ and its superspace $\mbox{met}(DT)$. The Wheeler--DeWitt for the functional $\Psi[DT(\{\ell\})]$ is easily obtained by the local one so that we have 
\begin{equation}
\label{dWdW}
\left\{-(16\pi G)^{2}\sum_{i,j}\frac{G_{ij}(\tau)}{4\ell_{i}\ell_{j}}\frac{\partial^{2}}{\partial \ell_{i}\partial \ell_{j}} -\sum_{h}\ell_{h}\beta_{h}+\Lambda V_{DT}\right\}\Psi[DT(\{\ell\})]=0 .
\end{equation}
Equal lengths Einstein metrics are indeed the natural reference metrics in this case because the numerical factor in front of the Regge action is exactly what is needed to satisfy the requirements of Def. \ref{eindef} for classical metrics to be stationary configurations for gravity with positive cosmological constant.

Quantum fluctuations with respect to any such classical configuration can be now parametrized in terms of the conformal invariants $\xi$ and $\eta$, and of the weights $\mathfrak{f}_{j}$  of the four vertices (see Fig. \ref{smpic}). 

A detailed analysis of the phase structure of these fluctuations, the semiclassical limit as well as explicit calculations will be presented elsewhere.

\section*{Acknowledgments}
We acknowledge partial support from PRIN 2010--11 \emph{Geometric and analytic theory of finite and infinite dimensional Hamiltonian systems}.

\appendix
\section{Simplicial metrics}
In each compact $d-$dimensional simplicial PL manifold a $d-$simplex $s$ can be endowed with a Euclidean metric, choosing as a coordinate basis the $d$ edges sharing a a vertex labeled $0$ in the following way (\cite{ham1})
\begin{equation}
g_{ij}(s)=\frac{1}{2}(\ell_{0i}^{2}+\ell_{0j}^{2}-\ell_{ij}^{2}).
\label{simpmet}
\end{equation}
It should be stressed that the indices of $g$ have a tensorial character, i.e. they indicate the components of the tensor $g$ in the basis 
$\{\ell_{0i}\}_{i=1,\cdots, d}$. On the other hand, the same indices on the length $\ell_{ij}$ are a mere label:  $\ell_{ij}$ indicates simply the length of the edge connecting the vertices $i$ and $j$ with, $i,j\in\{0,1,\dots,d\}$ and $i\neq j$.

This expression for the metric $g$ allows us to express the volume of a $d-$simplex $s$ in terms of its edge lengths:
\begin{equation}
V_{d}(s)=\frac{1}{d!}\sqrt{\det g_{ij}(s)}=\frac{1}{d!}\sqrt{\frac{(-1)^{d+1}}{2^{d}}\mbox{CM}_{d}(s)},
\end{equation}
where $\mbox{CM}_{d}$ is the \emph{Cayley-Menger determinant} in $d$ dimensions, i.e.
\begin{equation}
\mbox{CM}_{d}(s)=\det\left(\begin{array}{cccccc}
0 & 1 & 1 & 1 & \cdots & 1\\ 
1 & 0 & \ell^{2}_{01} & \ell^{2}_{02} & \cdots & \ell^{2}_{0d}\\ 
1 & \ell^{2}_{01} & 0 & \ell^{2}_{12} & \cdots & \ell^{2}_{1d}\\ 
1 & \ell^{2}_{02} & \ell^{2}_{12} & 0 & \cdots & \ell^{2}_{2d}\\
\vdots & \vdots & \vdots & \vdots & \ddots & \vdots\\ 
1 & \ell^{2}_{0d} & \ell^{2}_{1d} & \ell^{2}_{2d} & \cdots & 0
\end{array}\right)
\label{CM}
\end{equation}
Moreover we can obtain the following expression for dihedral angles 
\begin{equation}
\sin(\alpha_{\kappa\kappa'})=\frac{d}{d-1}\frac{V_{d}(s)V_{d-2}(h)}{V_{d-1}(\kappa)V_{d-1}(\kappa')},
\end{equation}
where $V_{d-2}(h)$ is the $(d-2)-$dimensional volume of the hinge $h$ and $V_{d-1}(\kappa)$ is the $(d-1)-$dimensional volume of the face $\kappa$. 
In three dimensions dihedral angles can also be explicitly expressed in terms of the edges of the tetrahedron (3-simplex) shown in Fig.\ref{labelt} as  
\begin{equation}
\label{dihang}
\cos{\alpha_{ij}}=\frac{2\left(\ell_{ik}^{2}+\ell_{il}^{2}-\ell_{kl}^{2}\right)\ell_{ij}^{2}-\left(\ell_{ij}^{2}+\ell_{ik}^{2}-\ell_{jk}^{2}\right)\left(\ell_{ij}^{2}+\ell_{il}^{2}-\ell_{jl}^{2}\right)}{\sqrt{\left[4\ell_{ij}^{2}\ell_{ik}^{2}-\left(\ell_{ij}^{2}+\ell_{ik}^{2}-\ell_{jk}^{2}\right)^{2}\right]\left[4\ell_{ij}^{2}\ell_{il}^{2}-\left(\ell_{ij}^{2}+\ell_{il}^{2}-\ell_{jl}^{2}\right)^{2}\right]}},
\end{equation}
where we have labeled  the edges in terms of vertices $i,j,k$ and $l$.
\begin{figure}[h!]
\begin{center}
\includegraphics[scale=0.3]{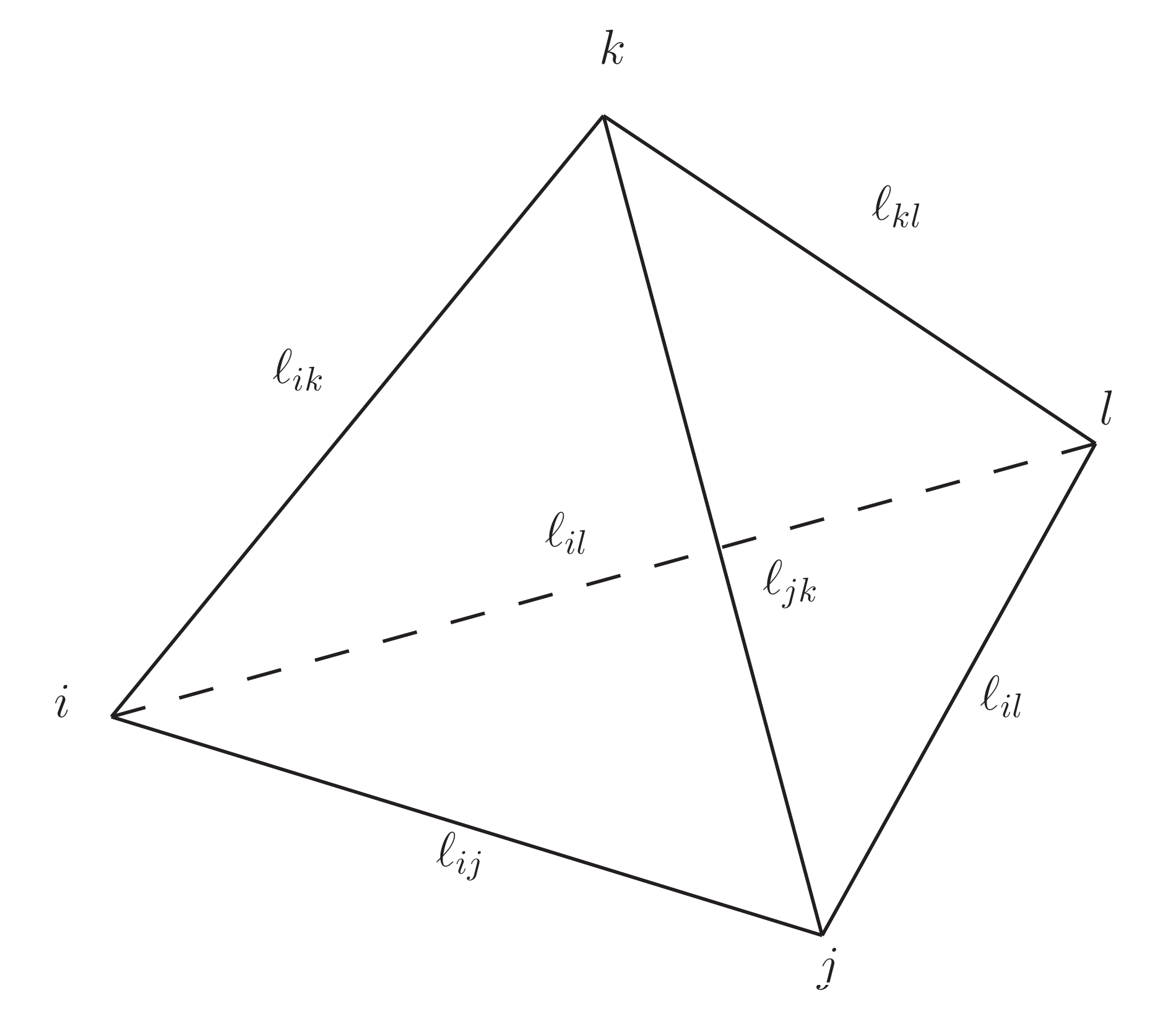}
\caption{\small{Labeling  of a tetrahedron}}
\label{labelt}
\end{center}
\end{figure}

\index{Bibliography}

\end{document}